\documentclass[10pt,conference]{IEEEtran}

\usepackage{fancyhdr}
\usepackage[normalem]{ulem}
\usepackage[hyphens]{url}
\usepackage[sort,nocompress]{cite}
\usepackage[final]{microtype}
\usepackage[keeplastbox]{flushend}
\usepackage[binary-units]{siunitx}
\usepackage{algorithm}
\usepackage{algorithmic}
\usepackage{multirow}
\usepackage{booktabs}
\usepackage{fancyhdr}
\usepackage{subcaption}
\usepackage{tikz}
\usepackage{soul} 

\usepackage{float}
\usepackage[bookmarks=true,breaklinks=true,letterpaper=true,colorlinks,citecolor=black,linkcolor=black,urlcolor=black]{hyperref}

\usepackage{xcolor}

\newcommand*\circled[1]{\tikz[baseline=(char.base)]{
            \node[shape=circle,fill,inner sep=1.5pt] (char) {\textcolor{white}{#1}};}}

\newcommand{\hpcayear}{2024}

\makeatletter
\newcommand{\linebreakand}{%
  \end{@IEEEauthorhalign}
  \hfill\mbox{}\par
  \mbox{}\hfill\begin{@IEEEauthorhalign}
}
\makeatother

\newcommand{\hpcasubmissionnumber}{8}
\title{A Two Level Neural Approach Combining Off-Chip Prediction with Adaptive Prefetch Filtering} 

\def\hpcacameraready{} 

\author{
    \ifdefined\hpcacameraready
        \IEEEauthorblockN{Alexandre Valentin Jamet\\
            \href{mailto:alexandre.jamet@bsc.es}{\texttt{alexandre.jamet@bsc.es}}
        }
        \IEEEauthorblockA{
            Barcelona Supercomputing Center (BSC)\\Universitat Politècnica de Catalunya (UPC)
        }
        \and
        \IEEEauthorblockN{Georgios Vavouliotis\\
            \href{mailto:georgios.vavouliotis2@huawei.com}{\texttt{georgios.vavouliotis2@huawei.com}}
        }
        \IEEEauthorblockA{
            Huawei Zurich Research Center
        }
        \and
        \IEEEauthorblockN{Daniel A. Jiménez\\
            \href{mailto:djimenez@acm.org}{\texttt{djimenez@acm.org}}
        }
        \IEEEauthorblockA{
            Texas A\&M University
        }
        \linebreakand
        \IEEEauthorblockN{Lluc Alvarez\\
            \href{mailto:lluc.alvarez@bsc.es}{\texttt{lluc.alvarez@bsc.es}}
        }
        \IEEEauthorblockA{
            Barcelona Supercomputing Center (BSC)\\Universitat Politècnica de Catalunya (UPC)
        }
        \and
        \IEEEauthorblockN{Marc Casas\\
            \href{mailto:marc.casas@bsc.es}{\texttt{marc.casas@bsc.es}}
        }
        \IEEEauthorblockA{
            Barcelona Supercomputing Center (BSC)\\Universitat Politècnica de Catalunya (UPC)
        }
    \else
        \IEEEauthorblockN{\normalsize{HPCA \hpcayear{} Submission
        \textbf{\#\hpcasubmissionnumber{}}} \\
        \IEEEauthorblockA{
            Confidential Draft \\
            Do NOT Distribute!!
        }
    }
    \fi
}

\def\aeopen{}           
\def\aereviewed{}     

\fancypagestyle{camerareadyfirstpage}{%
  \fancyhead{}
  
  \fancyhead[C]{
    \ifdefined\aeopen
    \parbox[][12mm][t]{13.5cm}{\hpcayear{} IEEE International Symposium on High-Performance Computer Architecture (HPCA)}    
    \else
      \ifdefined\aereviewed
      \parbox[][12mm][t]{13.5cm}{\hpcayear{} IEEE International Symposium on High-Performance Computer Architecture (HPCA)}
      \else
      \ifdefined\aereproduced
      \parbox[][12mm][t]{13.5cm}{\hpcayear{} IEEE International Symposium on High-Performance Computer Architecture (HPCA)}
      \else
      \parbox[][0mm][t]{13.5cm}{\hpcayear{} IEEE International Symposium on High-Performance Computer Architecture (HPCA)}
    \fi 
    \fi 
    \fi 
    \ifdefined\aeopen 
      \includegraphics[width=12mm,height=12mm]{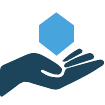}
    \fi 
    \ifdefined\aereviewed
      \includegraphics[width=12mm,height=12mm]{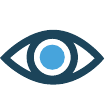}
    \fi 
    \ifdefined\aereproduced
      \includegraphics[width=12mm,height=12mm]{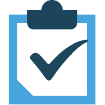}
    \fi
  }
  \fancyfoot[C]{}
}
\begin{document}
\maketitle

\ifdefined\hpcacameraready 
  \thispagestyle{camerareadyfirstpage}
  \pagestyle{empty}
\else
  \thispagestyle{plain}
  \pagestyle{plain}
\fi

\newcommand{\hpcaheight}{0mm}
\ifdefined\eaopen
\renewcommand{\hpcaheight}{12mm}
\fi


\begin{abstract}

To alleviate the performance and energy overheads of contemporary applications with large data footprints, we propose the \emph{Two Level Perceptron (TLP)} predictor, a neural mechanism that effectively combines predicting whether an access will be off-chip with adaptive prefetch filtering at the first-level data cache (L1D). 
TLP is composed of two connected microarchitectural perceptron predictors, named \emph{First Level Predictor (FLP)} and \emph{Second Level Predictor (SLP)}. 
FLP performs accurate off-chip prediction by using several program features based on virtual addresses and a novel selective delay component. 
The novelty of SLP relies on leveraging off-chip prediction to drive L1D prefetch filtering by using physical addresses and the FLP prediction as features.
TLP constitutes the first hardware proposal targeting both off-chip prediction and prefetch filtering using a multi-level perceptron hardware approach. TLP only requires 7KB of storage.

To demonstrate the benefits of TLP we compare its performance with state-of-the-art approaches using off-chip prediction and prefetch filtering on a wide range of single-core and multi-core workloads. 
Our experiments show that TLP reduces the average DRAM transactions by 30.7\% and 17.7\%, as compared to a baseline using state-of-the-art cache prefetchers but no off-chip prediction mechanism, across the single-core and multi-core workloads, respectively, while recent work significantly increases DRAM transactions.
As a result, TLP achieves geometric mean performance speedups of 6.2\% and 11.8\% across single-core and multi-core workloads, respectively. 
In addition, our evaluation demonstrates that TLP is effective independently of the L1D prefetching logic.

\end{abstract}

\section{Introduction}
\label{sec:introduction}

Emerging workloads from various domains \cite{Have13,basak_analysis_2019,10.1145/3159652.3159731,Brin98,Evelien02} have large data footprints that are orders of magnitude larger than the capacity of current cache hierarchies~\cite{10.1145/1454115.1454128}. These workloads frequently trigger DRAM accesses, spending a large portion of their execution time waiting for data transfers to and from DRAM to complete with a detrimental effect on performance and energy~\cite{zhang2017making,balaji2018graph,balaji2019combining, 10.1145/2925426.2926254, beamer2017reducing,10.1145/2882903.2915220}.  

Prior work has proposed several techniques to mitigate the performance and energy overheads of these applications. These techniques can be broadly classified into four categories: (i) off-chip prediction schemes that predict whether a memory access will result in a DRAM access or hit in the cache hierarchy \cite{jalili_locmap,10.1145/2678373.2665694,1183548, 10.1145/307338.300983,bera_hermes}, (ii) aggressive data prefetching with adaptive filters to ensure that only correct prefetches will be issued~\cite{bhatia2019perceptron},  (iii) cache bypassing that avoids caching blocks that will not be referenced in the near future~\cite{jaleel_high_2010,5695535,wu2011ship,7783705,jimenez_multiperspective_2017,shi_applying_2019}, and (iv) disruptive cache designs and optimizations for specific workload types~\cite{qureshi_line_2007,7551391,10.5555/2685048.2685096,8327036,8327035,7920847,ppa,7284059}. This work focuses on the first two categories and aims at combining their benefits in a cost-effective manner.

Despite their potential for determining the location of requested data in the memory hierarchy, previously proposed off-chip predictors \cite{bera_hermes,jalili_locmap} have important drawbacks that undermine their potential for boosting the performance of the memory subsystem while hindering their implementation in real-world designs. For example, the state-of-the-art off-chip predictor \cite{bera_hermes} triggers two memory accesses, one to DRAM and a second regular request to the cache hierarchy, when it predicts that the corresponding load access will be served from DRAM. While this approach can potentially reduce the latency of a load request that ends up being served from DRAM, it may also significantly increase the number of DRAM transactions. 
This work shows that, although effective, the state-of-the-art off-chip predictor significantly increases the number of DRAM transactions, which is a critical aspect in bandwidth-constrained scenarios. 
In addition, our analysis indicates that a large fraction of the inaccurate off-chip predictions is actually served by the first-level data cache (L1D). Therefore, a microarchitectural scheme that selectively delays the off-chip predictions with modest confidence until the L1D lookup is resolved has potential to significantly reduce the number of useless DRAM transactions and deliver higher performance.

Previous approaches have successfully applied prefetch filtering at the lower level caches~\cite{10.1145/2370816.2370868, 1240591, 10.1145/2677956,bhatia2019perceptron}. However, these approaches are not agile since they are typically optimized on top of specific prefetch engines, incur significant area overheads, and are not exposed to program features that are very valuable to produce accurate predictions ({\em e.g.}, a complete sequence of accessed virtual addresses). This work argues that the concept of off-chip prediction can be leveraged to form effective prefetch filters for L1D. Specifically, our analysis demonstrates that the vast majority of the L1D prefetch requests served from DRAM are inaccurate. 

To address our findings and improve the performance or memory-intensive workloads, we propose the \emph{Two Level Perceptron (TLP)} predictor. 
TLP constitutes the first hardware proposal targeting both off-chip prediction and prefetch filtering using a multi-level perceptron hardware approach. 
TLP is composed of two connected microarchitectural perceptron predictors: the \emph{First Level Predictor (FLP)} and the \emph{Second Level Predictor (SLP)}. FLP is a perceptron hardware predictor located near the core that employs a novel mechanism to reduce the number of DRAM accesses by selectively delaying off-chip predictions when needed. 
SLP is a perceptron predictor located alongside the L1D.
The novelty of SLP relies on leveraging off-chip prediction to drive L1D prefetch filtering using physical addresses as well as the FLP prediction as features.
Our evaluation illustrates that TLP yields significantly higher performance than the state-of-the-art off-chip predictor~\cite{bera_hermes} and prefetch filtering scheme~\cite{bhatia2019perceptron} 
across a large set of single-core and multi-core workloads.  

This paper makes the following contributions: 



\textbullet~~We design and propose \emph{Two Level Perceptron (TLP)} predictor, a scheme composed of two connected perceptron predictors: FLP and SLP. 
FLP reduces the pressure on the memory subsystem using a novel selective delay mechanism. 
SLP leverages off-chip prediction to guide prefetch filtering in the L1 data cache.
TLP is the first hardware proposal targeting both off-chip prediction and prefetch filtering.
TLP only requires 7KB of storage.

\textbullet~~We compare TLP with the state-of-the-art off-chip predictor, Hermes \cite{bera_hermes}, the state-of-the-art prefetch filter, PPF \cite{bhatia2019perceptron}, and a combination of both.
Our evaluation considers 55 single-core and 200 multi-core workloads. 
When considering a system that uses IPCP~\cite{9138971} as L1D prefetcher, 
TLP reduces the average number of DRAM transactions by 30.7\% and 17.7\%, as compared to a baseline that uses IPCP as L1D prefetcher but no off-chip prediction mechanism, across the single-core and multi-core workloads, respectively, while state-of-the-art approaches significantly increase DRAM transactions. 
As a result, TLP achieves geometric mean performance speedups of 6.2\% and 11.8\% across single-core and multi-core workloads, respectively. 
When considering a scenario with the Berti \cite{navarro2022berti} L1D prefetcher, TLP also outperforms Hermes, PPF, and a combination of them in both single-core and multi-core contexts since it significantly reduces DRAM accesses. 


\section{Background} 
\label{sec:background_motivation}



\subsection{Off-Chip Prediction}
\label{subsec:offchip_prediction}


Emerging workloads spanning various domains \cite{10.1145/3159652.3159731,Have13,basak_analysis_2019,Brin98,Evelien02}, have a key property in common: massive working set sizes that do not fit in the existing cache hierarchies \cite{10.1145/1454115.1454128}, making cache management a major performance bottleneck for processor design. Indeed, recent work~\cite{zhang2017making, balaji2018graph, balaji2019combining, 10.1145/2925426.2926254, beamer2017reducing, 10.1145/2882903.2915220} shows that these workloads spend up to 80\% of their total execution time waiting for DRAM. 

To address the high-latency load requests of these emerging applications, prior work~\cite{jalili_locmap, 10.1145/2678373.2665694, 1183548, 10.1145/307338.300983, bera_hermes} has introduced the concept of \emph{off-chip prediction}. The core idea behind off-chip prediction is to predict whether a memory access will eventually result in a DRAM access or in a hit in the cache hierarchy (L1D, L2C, LLC). Prior work in the domain can be classified in two categories depending on their prediction strategy: i) predict which cache level (L1D, L2C, LLC) will provide a hit, if any \cite{10.1145/2678373.2665694,jalili_locmap}, and ii) predict whether the cache hierarchy as a whole will provide a hit or not \cite{bera_hermes}. A representative work from the first category is Level Prediction (LP) \cite{jalili_locmap}, a scheme that dynamically predicts where in the memory hierarchy a demanded memory block is most likely to be found. A representative scheme of the second category is Hermes~\cite{bera_hermes}, an adaptive perceptron-based off-chip predictor that routes demand load requests directly to DRAM when it is confident that the load will miss in all cache levels.


Hermes~\cite{bera_hermes} is the state-of-the-art microarchitectural off-chip prediction scheme. At the core of Hermes, there is a perceptron predictor composed of several prediction tables, one per selected program feature, similar to prior work on perceptron-based microarchitectural prediction: from branch prediction~\cite{garza2019bit,1431572,903263} to cache replacement policies~\cite{7783705, jimenez_multiperspective_2017} and other intelligent modules~\cite{bhatia2019perceptron}. 

Hermes is consulted to provide a prediction upon demand load requests. If the prediction is positive (\emph{i.e.}, the demand load request is predicted to go off-chip), the core issues two requests: one regular request to the cache hierarchy, that might go down to DRAM, and another speculative request that fetches the cache line from DRAM in an attempt to hide the latency cost of accessing the caches.
When a demand request eventually returns to the core to be consumed, the training logic of Hermes compares the original prediction with the actual outcome and accordingly updates the weights in the prediction tables. 



\subsection{Prefetch Filtering}
\label{subsec:prefetch_filtering}

Hardware prefetching is a technique that proactively fetches blocks in the cache hierarchy before they are explicitly requested by a core. 
Hardware prefetchers need to deal with two metrics that are at odds with one another: miss coverage and prefetching accuracy. Aggressive prefetchers typically have high coverage but low accuracy while conservative prefetchers tend to have low coverage and high accuracy.

To handle the coverage-accuracy trade-off, smart prefetch filters and throttling schemes able to accurately identify useless prefetch requests and discard them have been proposed \cite{10.1145/2370816.2370868, 1240591, 10.1145/2677956,bhatia2019perceptron}. An effective prefetch filter would increase the accuracy of a hardware prefetcher without harming its coverage, resulting in higher performance by enabling better cache management. 

The state-of-the-art prefetch filter is the Perceptron-based Prefetch Filter (PPF) \cite{bhatia2019perceptron}, a perceptron predictor that uses several program features to filter out inaccurate prefetch requests, increasing the accuracy of the underlying prefetcher. Although effective, PPF has two limitations. First, PPF is built and optimized on top of a specific prior prefetcher \cite{kim2016path}, thus it requires significant engineering effort as well as feature exploration and tuning to make it filter effectively the requests of other prefetchers. Second, PPF incurs 40KB of storage overhead, which hinders its adoption by commercial designs.

\section{Motivation}
\label{sec:motivation}



This section motivates the need for better off-chip predictors and highlights the potential of leveraging the concept of off-chip prediction to apply effective prefetch filtering for the L1D cache.
Section \ref{subsec:cache_behavior} characterizes the cache behavior of contemporary applications, showing that a large fraction of the memory accesses that miss in the L1D result in a DRAM access. 
Section \ref{subsec:impact_hermes} analyzes the behavior of Hermes~\cite{bera_hermes}, the state-of-the-art off-chip predictor presented in Section \ref{subsec:offchip_prediction}, in both single-core and multi-core contexts. 
Our analysis indicates that Hermes significantly increases DRAM bandwidth consumption, especially in multi-core contexts. 
Therefore, performance improvements are possible by reducing the number of additional DRAM transactions triggered by Hermes. 
Section~\ref{subsec:off_chip_prediction_for_l1d_prefetch_filtering} focuses on L1D cache prefetching and characterizes the inaccurate prefetches issued by two state-of-the-art L1D prefetchers,
and reveals that off-chip prediction can drive the design of effective prefetch filters for L1D. 
%
%
Section~\ref{sec:methodology} presents in detail our simulation infrastructure and all the considered workloads.

\begin{figure}
    \centering
    \includegraphics[width=\columnwidth]{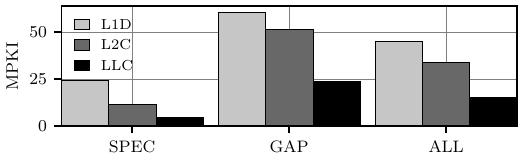}
    \caption{MPKI of all caches (L1D, L2C, LLC) across the SPEC (SPEC CPU 2006 and SPEC CPU 2017) and GAP workloads.}
    \vspace{-0.5cm}
    \label{fig:baseline_mpkis}
\end{figure}

\subsection{Cache Behavior of Modern Workloads}
\label{subsec:cache_behavior}

Prior work discussed in Section \ref{subsec:offchip_prediction} shows that the majority of demand load requests of applications featuring huge data working sets miss in all levels of the cache hierarchy, triggering many DRAM accesses. This section analyzes the cache behavior of all single-core workloads presented in Section~\ref{sec:methodology}. 


Figure~\ref{fig:baseline_mpkis} shows the average Misses per Kilo Instruction (MPKI) rates of L1D, L2C and LLC. On average the MPKIs of L1D, L2C, and LLC are 45.0, 34.1, and 15.6, respectively. Therefore, 34.7\% of L1D misses eventually require a DRAM access. Remarkably, workloads from domains such as graph processing put more pressure on the cache hierarchy, resulting in more frequent DRAM accesses. Indeed, Figure \ref{fig:baseline_mpkis} reveals that, on average, the graph-processing (GAP) workloads trigger a DRAM access for 39.7\% of the L1D misses.

\vspace{0.3cm}
\noindent\fcolorbox{black}{gray!10}{
    \parbox{\dimexpr\linewidth-2\fboxsep-2\fboxrule-7pt}{
        \emph{\textbf{Finding 1.} 
            A large fraction of the demand load requests triggered by applications with large working set sizes miss in all cache levels. 
        } 
    }
}
\vspace{0.2cm}

\subsection{Impact of Hermes}
\label{subsec:impact_hermes}

\begin{figure}
    \centering
    \includegraphics[width=\columnwidth]{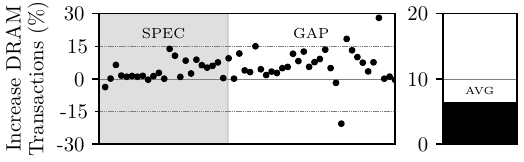}
    \caption{Increase in DRAM transactions due to Hermes off-chip predictions relative to a baseline without off-chip prediction mechanism. 
    Lower is better.}
    \vspace{-0.5cm}
    \label{fig:hermes_single_core_motivation}
\end{figure}

\begin{figure}
    \centering
    \includegraphics[width=\columnwidth]{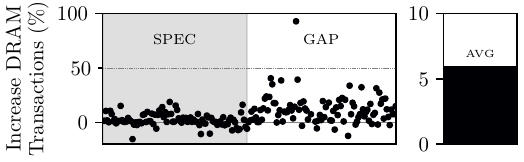}
    \caption{Increase in DRAM transactions due to Hermes off-chip predictions relative to a baseline without off-chip prediction mechanism in the 4-core context. The x-axis ticks represent 200 different 4-core workload mixes of SPEC and GAP workloads. 
    Lower is better. 
    }
 \vspace{-0.5cm}
    \label{fig:hermes_multi_core_motivation}
\end{figure}

This section quantifies the impact of Hermes on the number of DRAM transactions processed by the main memory in both single-core and multi-core contexts and identifies features that can potentially increase Hermes' efficiency and performance. This analysis is conducted using the methodology and the set of workloads presented in Section~\ref{sec:methodology}.

\subsubsection{DRAM Transactions} 
\label{subsubsec:dram_transactions}

Figures \ref{fig:hermes_single_core_motivation} and \ref{fig:hermes_multi_core_motivation} illustrate the impact of Hermes on the number of DRAM transactions in single-core and multi-core contexts, respectively. 
The x-axis display different SPEC and GAP workloads.
Both SPEC and and GAP workloads are separately sorted considering the LLC MPKI.
The y-axis displays the increase in terms of DRAM transactions that Hermes incurs  over a baseline without any off-chip predictor.

 Figures~\ref{fig:hermes_single_core_motivation} and~\ref{fig:hermes_multi_core_motivation} indicate that Hermes places high pressure on DRAM, especially in the multi-core scenario, since it issues many speculative DRAM requests. Regarding the single-core evaluation, Hermes increases the number of DRAM transactions by 5.2\%, 6.6\%, and 6.4\% over the baseline system that does not use any off-chip predictor for the SPEC, GAP, and all workloads combined, respectively. 
 Figure \ref{fig:hermes_multi_core_motivation}, which presents the impact of Hermes on DRAM transactions in a multi-core context, shows that Hermes significantly increases DRAM transactions. Specifically, Hermes increases the average number of DRAM transactions by 2.2\%, 9.6\%, and 6.0\% over the multi-core baseline for the SPEC mixes, GAP mixes, and all mixes, respectively. Notably, the increase in DRAM transactions for the GAP workloads is significantly higher than the increase for the SPEC workloads; this happens because the GAP suite is made of graph-processing applications that have much larger data working sets than the general-purpose SPEC CPU workloads. 

\vspace{0.3cm}

\noindent\fcolorbox{black}{gray!10}{
    \parbox{\dimexpr\linewidth-2\fboxsep-2\fboxrule-7pt}{
        \emph{\textbf{Finding 2.} Hermes significantly increases the number of DRAM accesses in both single-core and multi-core contexts, especially for graph-processing applications.
        }
    }
}
\vspace{0.2cm}

\subsubsection{Analysis of Hermes Predictions} 
\label{subsubsec:analysis_of_hermes_predictions}

\begin{figure}
    \centering
    \includegraphics{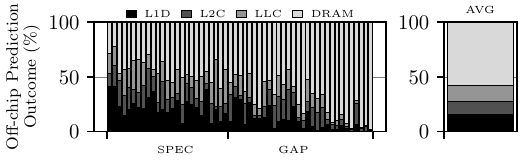}
    \caption{
    Location of a block upon a Hermes off-chip prediction. 
    }
    \vspace{-0.7cm}
    \label{fig:hermes_offchip_pred_l1d_hit}
\end{figure}

This section characterizes the off-chip predictions of Hermes (\emph{i.e.}, cases where Hermes triggered a speculative DRAM request), 
and motivates potential design and functionality enhancements. To do so, we categorize the off-chip predictions of Hermes depending on where the corresponding block is located in the memory hierarchy (L1D, L2C, LLC, DRAM). Specifically, we consider the following categories: (i) block resides in L1D, (ii) block resides in L2C, (iii) block resides in LLC, and (iv) block resides in DRAM. Predictions belonging to categories (i), (ii), and (iii) correspond to inaccurate off-chip predictions since the block is located in the cache hierarchy while category (iv) represents accurate off-chip predictions since the block is not present in the caches. Figure~\ref{fig:hermes_offchip_pred_l1d_hit} presents this breakdown for both SPEC and GAP single-core workloads, using the methodology that Section \ref{sec:methodology} describes. Both SPEC and GAP workloads are separately sorted based on LLC MPKI, similar to Figure \ref{fig:hermes_single_core_motivation}.

Figure \ref{fig:hermes_offchip_pred_l1d_hit} shows that 42.2\% of the total off-chip predictions are inaccurate since the corresponding blocks reside in the cache hierarchy (L1D , L2C, or LLC). Notably, a large fraction of the load requests corresponding to an inaccurate off-chip prediction are served by the L1D cache. Specifically, 17.7\% of the total off-chip predictions are useless since their corresponding block resides in the L1D. In other words, delaying Hermes to issue an off-chip prediction after the L1D lookup completion would significantly reduce DRAM transactions. However, constantly delaying the off-chip predictions of Hermes until the L1D lookup is completed would result in suboptimal performance gains since more than 50\% (57.8\% on average in Figure \ref{fig:hermes_offchip_pred_l1d_hit}) of the Hermes off-chip predictions are accurate. In these cases, issuing the DRAM access before the L1D access is resolved provides latency benefits. Thus, a mechanism to decide whether or not an off-chip prediction of Hermes should be issued before or after the L1D access completion has the potential to significantly reduce the number of useless DRAM accesses triggered by Hermes.

\vspace{0.3cm}
\noindent\fcolorbox{black}{gray!10}{
    \parbox{\dimexpr\linewidth-2\fboxsep-2\fboxrule-7pt}{
        \emph{\textbf{Finding 3.} 
            Selectively delaying Hermes off-chip predictions until the L1D lookup is resolved has the potential to significantly reduce the number of useless DRAM transactions and deliver higher performance.
        }
    }
}
\vspace{0.2cm}

\subsection{Off-Chip Prediction for L1D Prefetch Filtering}
\label{subsec:off_chip_prediction_for_l1d_prefetch_filtering}



\begin{figure}
    \centering
    \begin{subfigure}[b]{\columnwidth}
        \centering
        \includegraphics{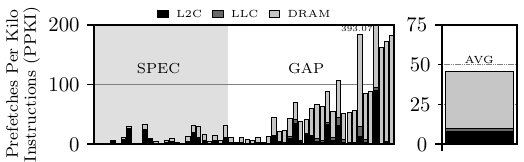}
        \caption{IPCP}
        \label{fig:l1d_ipcp_useless_prefetches_breakdown}
    \end{subfigure}
    \hfill
    \begin{subfigure}[b]{\columnwidth}
        \centering
        \includegraphics{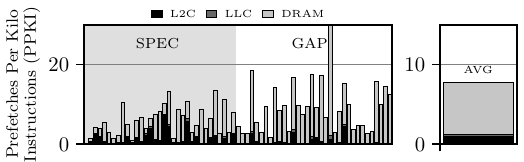}
        \caption{Berti}
        \label{fig:l1d_berti_useless_prefetches_breakdown}
    \end{subfigure}
    \caption{Location where the inaccurate L1D prefetch requests are served across two state-of-the-art L1D prefetchers. Both SPEC and GAP workloads are separately sorted based on LLC MPKI, similar to Figure \ref{fig:hermes_single_core_motivation}. 
    }
    \label{fig:l1d_useless_prefetches_breakdown}
\end{figure}

\begin{figure}
    \centering
    \begin{subfigure}[b]{\columnwidth}
        \centering
        \includegraphics{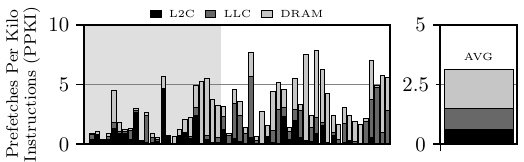}
        \caption{IPCP
        }
        \label{fig:l1d_ipcp_useful_prefetches_breakdown}
    \end{subfigure}
    \hfill
    \begin{subfigure}[b]{\columnwidth}
        \centering
        \includegraphics{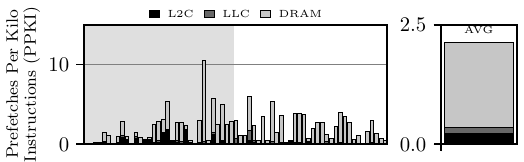}
        \caption{Berti
        }
        \label{fig:l1d_berti_useful_prefetches_breakdown}
    \end{subfigure}
    \caption{
    Location where the accurate L1D prefetch requests are served across two state-of-the-art L1D prefetchers. Both SPEC and GAP workloads are separately sorted based on LLC MPKI, similar to Figure \ref{fig:hermes_single_core_motivation}.
    }
    \vspace{-0.5cm}
    \label{fig:l1d_useful_prefetches_breakdown}
\end{figure}

This section characterizes the inaccurate prefetches issued by L1D prefetchers across the considered single-core SPEC and GAP workloads. To do so, we consider two state-of-the-art L1D prefetchers: (i) the Instruction Pointer Classification Prefetcher (IPCP) \cite{9138971}, and (ii) the Berti prefetcher \cite{navarro2022berti}.

Figure \ref{fig:l1d_ipcp_useless_prefetches_breakdown} presents the breakdown of the inaccurate L1D prefetches issued by IPCP depending on where in the memory hierarchy (L2C, LLC, DRAM) the corresponding prefetch request is served. To do so, we use the Prefetches Per Kilo Instruction (PPKI) metric. Overall, 18.2\%, 3.8\%, and 78\% of the total inaccurate prefetch requests are served by L2C, LLC, and DRAM, respectively. We observe that the majority of the inaccurate prefetch requests are the ones that were served from DRAM. This behavior is more prevalent for the GAP workloads since these workloads have more complex patterns than SPEC. In addition, we compare the accurate L1D prefetches of IPCP that were served from DRAM with the inaccurate ones. Our analysis indicates that, on average, 95.2\% of the prefetches that were served from DRAM are inaccurate (the rest 4.8\% is accurate prefetches) for our set of workloads; the ratio is higher for the GAP benchmarks (96.7\%) than for the SPEC (82\%) since the former exhibit more complex memory access patterns. 
Figure~\ref{fig:l1d_berti_useless_prefetches_breakdown} presents the breakdown of Berti's inaccurate prefetches depending on where in the memory hierarchy the corresponding request is served, similar to Figure~\ref{fig:l1d_ipcp_useless_prefetches_breakdown}. Overall, we observe the same behavior. The vast majority of Berti´s useless prefetch requests are served from DRAM and there is high probability for a prefetch that goes all the way to DRAM to fetch a block to be inaccurate, making a strong case for exploiting the off-chip prediction technique to design an L1D prefetch filter.
Figure~\ref{fig:l1d_useful_prefetches_breakdown} indicates how the overall number of accurate prefetchers served from DRAM (1.7 and 0.5 PPKI for IPCP and Berti) is much smaller than the inaccurate prefetchers served from DRAM (35.6 and 6.6 for IPCP and Berti).

Thus, we conclude that accurately predicting whether a prefetch will be served from DRAM can provide a useful hint regarding the usefulness of the corresponding prefetch. Consequently, an accurate off-chip predictor can be leveraged as an L1D prefetch filter. Finally, we observe similar behavior in the multi-core context.

\vspace{0.3cm}
\noindent\fcolorbox{black}{gray!10}{
    \parbox{\dimexpr\linewidth-2\fboxsep-2\fboxrule\relax}{%
        \emph{\textbf{Finding 4.} 
            Off-chip prediction can be leveraged to design an effective prefetch filtering scheme for L1D.            
        }
    }%
}
\vspace{0.1cm}


These four findings demonstrate that the state-of-the-art approach for off-chip prediction incurs a significant overhead in terms of additional DRAM transactions, and that there are opportunities to eliminate this overhead and boost performance by unifying off-chip prediction and prefetch filtering.
Section~\ref{sec:two_level_perceptron} presents a novel approach that unifies these two techniques in a single method.

\section{Two Level Perceptron Prediction}
\label{sec:two_level_perceptron}

This paper proposes the \emph{Two Level Perceptron (TLP)} predictor, a two level cooperative prediction scheme that leverages neural methods to perform cost-effective off-chip prediction for demand load requests combined with adaptive L1D prefetch filtering. TLP is composed of two microarchitectural perceptron predictors named First Level Perceptron (FLP) predictor and Second Level Perceptron (SLP), respectively. 
TLP is motivated by the four findings of Section~\ref{sec:background_motivation}.

Findings 1 and 2 demonstrate that Hermes exacerbates the pressure on the memory subsystem that modern workloads inject, particularly for memory intensive workloads from domains like graph-processing.
In addition, Finding 3 indicates that a selective delay mechanism can potentially mitigate this pressure. 
The FLP design, described in Section~\ref{subsec:FLP_predictor}, is motivated by these three findings. FLP includes a novel selective delay mechanism to only trigger speculative requests to DRAM for highly confident off-chip predictions.
Finding 4 indicates the potential of guiding prefetch filtering via off-chip prediction.
The SLP design, described in Section~\ref{subsec:SLP_predictor}, exploits this potential and incorporates a novel feature based on FLP output.
Section~\ref{subsec:building_multi_Level_perceptrons} presents our complete proposal, TLP, a multi-Level perceptron combining FLP and SLP.
TLP is novel in three ways: i) it incorporates a new selective delay mechanism to reduce pressure on the memory subsytem; ii) it leverages off-chip prediction to guide prefetch filtering; and iii) it constitutes the first hardware proposal targeting both off-chip prediction and prefetch filtering.
In addition, TLP is the first multi-level perceptron hardware approach that can be effectively applied due to its low area requirements. Finally, Section~\ref{subsec:hardware_requirements} details the hardware requirements of TLP.

\subsection{First Level Perceptron (FLP) Predictor}
\label{subsec:FLP_predictor}

\begin{table}[]
    \begin{center}
        \small
        \setlength\tabcolsep{6pt}
        \bgroup
        \def\arraystretch{0.7}
        \resizebox{\columnwidth}{!}{%
        \begin{tabular}{lm{5cm}}
            \hline
                              &  \\
            \textbf{\textit{Legacy Hermes features}} & \begin{itemize}
                \item PC $\oplus$ cacheline offset
                \item PC $\oplus$ byte offset
                \item PC + first access
                    \item Cacheline offset + first access
                    \item Last-4 load PCs
                \end{itemize} \\
                \textbf{\textit{Leveling feature}} & \begin{itemize}
                    \item FLP prediction + cacheline offset
                \end{itemize} \\ \hline
        \end{tabular}
        }
        \egroup
    \end{center}
    \caption{List of features used by the FLP and the SLP.}
    \label{tab:flp_slp_features}
\end{table}

\begin{figure}
    \centering
    \includegraphics[width=0.8\columnwidth]{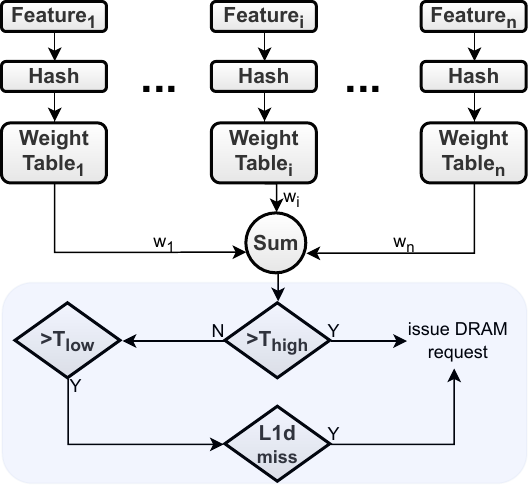}
    \caption{Flowchart of FLP. Diamonds indicate decision points.}
    \vspace{-0.5cm}
    \label{fig:flp_overview_flowchart}
\end{figure}

FLP is an off-chip predictor based on a micro-architectural hashed perceptron predictor that dynamically decides whether to consume the off-chip prediction in the core (\textit{i.e.}, in parallel with the L1D lookup since L1D caches are typically implemented as VIPT structures), or upon an L1D miss.
This delayed decision mechanism is driven by two threshold values: $\tau_{high}$ and $\tau_{low}$.
Perceptron confidence values greater than $\tau_{high}$ indicate a high probability for the corresponding load request to miss in all cache levels, values lower than  $\tau_{low}$ indicate the opposite, and intermediate values indicate the need for delaying the decision upon an L1D miss.
FLP takes into account several program features to predict whether a demand load request will miss in the cache hierarchy or not. Our exploration indicates that the features used in the original Hermes~\cite{bera_hermes} work provide good predictions and that adding more features provides marginal benefits. Thus, FLP uses the same set of features as the original Hermes prediction, presented in Table~\ref{tab:flp_slp_features} (\textit{c.f.}: Legacy Hermes features). The selected features correlate the probability of a demand load request going off-chip with a history of PCs and accessed memory regions. Each FLP feature is associated with a weight table which is composed of confidence counters. 

Figure~\ref{fig:flp_overview_flowchart} presents a flowchart of FLP's operation and illustrates how the confidence value produced by FLP is used to drive the off-chip prediction mechanism. Upon a demand load request, FLP is consulted by the core. FLP uses the selected program features to index its weight tables, then reads out and sums the corresponding weights to produce a confidence value. Then, the confidence value is compared to the $\tau_{high}$ threshold. A confidence value greater than $\tau_{high}$ indicates a high probability for the corresponding load request to miss in all caches. In this case, FLP issues a speculative DRAM request from the core 
in parallel with the L1D lookup as first-level caches are typically implemented as VIPT structures.
However, if the confidence value does not exceed $\tau_{high}$ but does exceed the $\tau_{low}$ threshold, the probability of the load demand request to miss in all cache levels is not considered high enough to benefit from a speculative DRAM request. Thus, the request is flagged as predicted off-chip and is sent to the L1D cache.
In Section~\ref{subsec:cache_behavior}, we observed that the probability of a load demand requiring an access to the DRAM tends to rise with each successive cache level traversed. Therefore, if this request results in an L1D miss, the flag bit is read, and a speculative DRAM request is issued from the L1D. Thus, FLP addresses our third analysis finding and avoids sending useless DRAM requests for loads that might hit in the on-chip caches. Finally, if the confidence value exceeds none of the two thresholds, the demand load request continues like a normal request without triggering speculative DRAM access.

The FLP is trained upon completing a memory access, (\textit{i.e.,} when the memory block is returned to the core from the cache hierarchy). When the request comes back to the core, the FLP checks if the request was a true off-chip load request (\textit{i.e.}, if this request required a DRAM access). If the request was a true off-chip load request, the predictor's corresponding weights are trained positively. Conversely, if the request was not a true off-chip load request, the predictor's corresponding weights are trained negatively. 

\subsection{Second Level Perceptron (SLP) Predictor}
\label{subsec:SLP_predictor}

\begin{figure}
    \centering
    \includegraphics[width=0.8\columnwidth]{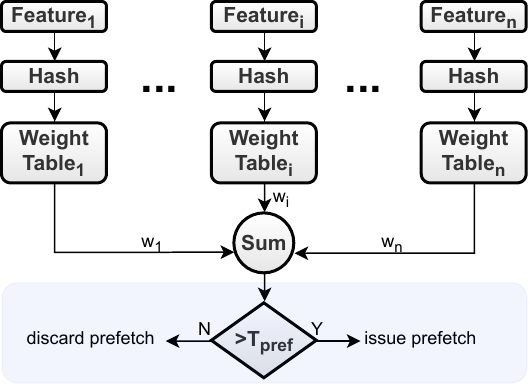}
    \caption{Flowchart of SLP. Diamonds indicate decision points.}
    \vspace{-0.5cm}
    \label{fig:slp_overview_flowchart}
\end{figure}

\begin{figure*}[h]
    \centering
    \includegraphics[width=\textwidth]{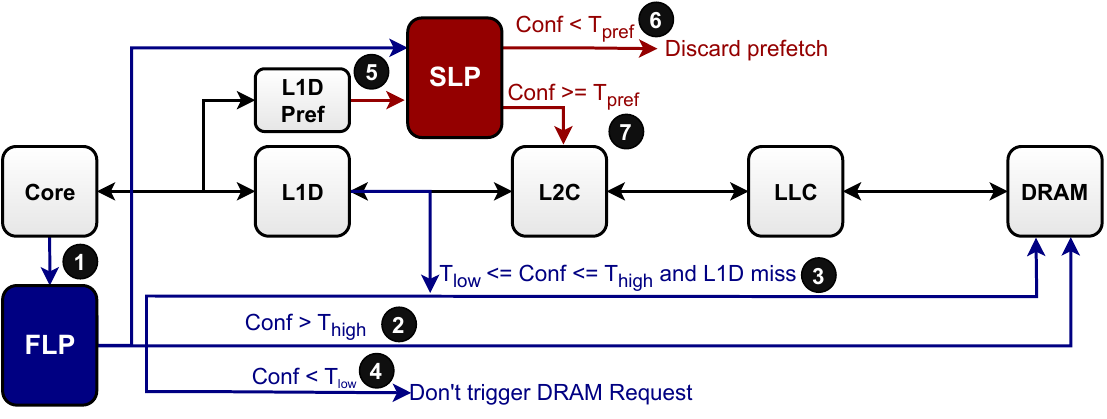}
    \caption{Organization and operation of the Two Level Perceptron (TLP) prediction mechanism.}
    \label{fig:tlp_overview}
\end{figure*}


The SLP is a perceptron-based off-chip predictor conceived to be used in the context of L1D prefetch filtering. The SLP design is motivated by the observation that off-chip prediction can be leveraged to design effective L1D prefetch filters. 
Section~\ref{subsec:off_chip_prediction_for_l1d_prefetch_filtering} justifies this observation.
SLP can be used to improve the performance of any generic L1D prefetcher since it makes no assumption regarding the L1D prefetcher design. 

SLP uses several program features to perform effective prefetch filtering at L1D. Our feature exploration indicates that FLP's features can be also used in the context of L1D prefetch filtering. Therefore, SLP uses the same as FLP, presented in Table \ref{tab:flp_slp_features}, but these features are adapted to use physical addresses in place of virtual addresses as SLP is placed after the L1D cache. Additionally, SLP makes use of a new feature denoted as FLP prediction + offset in Table \ref{tab:flp_slp_features}. This feature combines the FLP output bit of the cache block from which the prefetch request originated with the offset of the prefetched cache block in its physical memory page. 
The rationale of this feature is to correlate the probability of an L1D prefetch request going off-chip when a certain cache line offset is touched with the off-chip prediction decision related to the block that triggered the prefetch request. 
The SLP produces a binary off-chip prediction when an L1D prefetch request is issued.

Figure~\ref{fig:slp_overview_flowchart} presents a flowchart of the SLP operation. SLP is consulted when the L1D prefetcher issues a prefetch request. The confidence value is built similarly to the FLP. The output value is compared to the $\tau_{pref}$ threshold. If it exceeds $\tau_{pref}$, the prefetch is considered as eventually requiring a DRAM access and, therefore, likely useless. In this situation, the prefetch request is discarded. Conversely, if the confidence value does not exceed $\tau_{pref}$, the prefetch request is processed as usual by the cache hierarchy.

SLP is trained in a similar way as FLP (cf. Section~\ref{subsec:FLP_predictor}). Upon the completion of an L1D prefetch request, the predictor's weights are trained positively or negatively depending on whether or not the prefetch request was served off-chip.

\subsection{Building a Multi-Level Perceptron Predictor}
\label{subsec:building_multi_Level_perceptrons}

This section presents our complete proposal, Two Level Perceptron (TLP) predictor, a hierarchical neural prediction scheme that combines FLP and SLP predictors, presented in Sections \ref{subsec:FLP_predictor} and \ref{subsec:SLP_predictor}, respectively. 

Figure~\ref{fig:tlp_overview} shows the design and the operation of TLP. Upon a load demand access, the core consults FLP to obtain a confidence value $Conf$ driving the off-chip prediction \circled{1}. This prediction can give one of the three following outcomes: (i) the load request is predicted to be off-chip with high confidence ($Conf>\tau_{high}$), thus a speculative DRAM request is thrown from the core \circled{2} besides the regular load demand access; (ii) the load request is predicted to be off-chip with low confidence ($\tau_{low} \leq Conf \leq \tau_{high}$), thus the speculative DRAM request will be thrown only if the load misses in the L1D \circled{3}; (iii) the load request is predicted to be on-chip; therefore no additional action is taken \circled{4} besides triggering the regular demand access. Metadata relative to the prediction (hashed PC, history of last load PCs, and perceptron confidence value) are stored in the matching Load Queue entry for later training and an off-chip prediction tag is set in the load request thrown to the cache hierarchy depending on the FLP prediction. 

SLP is consulted upon L1D prefetch requests \circled{5}. To make a prediction, SLP takes as input the metadata attached to the prefetch request and the off-chip prediction tag attached to the demand load request from which the prefetch request originates. This information is used to produce an off-chip $Conf$ prediction specific to L1D prefetch request. This prediction can result in two possible outcomes: i) the prefetch request is predicted to be off-chip ($Conf < \tau_{pref}$) and the prefetch request is discarded \circled{6}, and ii) the prefetch request is predicted to be on-chip ($Conf \geq \tau_{pref}$) and the prefetch request is processed as usual by the cache hierarchy \circled{7}. Similarly to FLP, SLP stores metadata relative to its prediction in the L1D MSRH entries for later training.

The training routines of the FLP and the SLP are triggered upon completion of the corresponding requests, (\textit{i.e.}, for FLP when the load request returns to the core and SLP when the prefetch request is served), as Sections \ref{subsec:FLP_predictor} and \ref{subsec:SLP_predictor} explain.

\subsection{TLP Hardware Requirements and Latency}
\label{subsec:hardware_requirements}


Table~\ref{tab:storage_overhead} breaks down the hardware requirements of TLP into its various components. TLP only requires 6.98KB of additional storage per core. Similar to Hermes~\cite{bera_hermes}, FLP requires 3.21KB of storage for its prediction tables and 0.42KB of storage for the metadata in the Load Queue entries for training purposes. SLP requires 3.29KB of storage for its prediction tables as we make the addition of a new feature, and 0.06KB of additional storage in the L1D MSHR entries for training purposes. In total, TLP requires only 6.99KB of extra storage, making it a low overhead design combining off-chip prediction and prefetch filtering.
Similarly to previous work~\cite{bera_hermes}, we consider a 6-cycles latency when either FLP or SLP trigger a speculative DRAM access.

\begin{table}
    \begin{center}
        \small
        \setlength\tabcolsep{6pt}
        \bgroup
        \def\arraystretch{0.7}
        \resizebox{\columnwidth}{!}{%
        \begin{tabular}{lm{5cm}c}
            \hline
            Component & Description & Size \\
            \hline
                             &  \\
            \textbf{\textit{FLP}} & \begin{itemize}
                    \item Perceptron weight tables: 2.58KB
                    \item Page buffer: 0.63KB
                \end{itemize} & 3.21KB \\
                \textbf{\textit{SLP}} & \begin{itemize}
                    \item Perceptron weight tables: 2.66KB
                    \item Page buffer: 0.63KB
                \end{itemize} & 3.29KB \\ 
                \textbf{\textit{Load Queue metadata}} & Hashed PC: 32b; Last-4 PC: 10b; First access: 1b; perceptron confidence value: 5b & 0.42KB \\ 
                \textbf{\textit{L1D MSHR metadata}} & Hashed PC: 32b; Last-4 PC: 10b; First access: 1b; perceptron confidence value: 5b; prediction: 1b & 0.06KB \\ \hline
                \textbf{\textit{Total}} & & 6.98KB \\ \hline
        \end{tabular}
        }
        \egroup
    \end{center}
    \vspace{-0.3cm}
    \caption{Storage overhead of TLP.}
    \vspace{-0.5cm}
    \label{tab:storage_overhead}
\end{table}

\section{Experimental Methodology}
\label{sec:methodology}


\subsection{Simulation Methodology}
\label{subsec:simulation_infrastructure}

We evaluate our proposal using ChampSim \cite{gober2022championship}, a detailed trace-based simulator that models a 4-wide out-of-order CPU. We consider a baseline system similar to the Intel Cascade Lake microarchitecture~\cite{noauthor_cascade_nodate}. Table~\ref{table:cpu_config} presents the specific configuration details of the baseline system. 
Regarding hardware prefetching, we use state-of-the-art prefetchers in both the L1D and the L2C. At the L1D level we consider both the Instruction Pointer Classification Prefetcher (IPCP) prefetcher~\cite{9138971} and the Berti prefetcher~\cite{navarro2022berti}. At the L2 level we use the SPP prefetcher~\cite{kim2016path}, which brings prefetched blocks into either the L2C or the LLC depending on its internal prefetch logic. 

\subsection{Workloads}
\label{subsec:workloads}

Our evaluation considers a large set of applications spanning different benchmark suites. Specifically, we consider workloads from SPEC CPU 2006~\cite{spec2006} and SPEC CPU 2017~\cite{spec2017} benchmark suites. In addition, we consider graph-processing applications included in the GAP benchmark suite \cite{beamer_gap_2015}. Specifically, we use six graph-processing kernels from GAP: Breadth-First Search (BFS) is a fundamental graph traversal algorithm; Page Rank (PR) iteratively updates per-vertex ranks until convergence; Connected Components (CC) applies the Shiloach-Vishkin~\cite{shiloach1980log} algorithm to compute the largest connected components of the graph;  Betweenness Centrality (BC) uses the Brandes algorithm~\cite{brandes2001faster} to approximate the per-vertex centrality scores; Triangle Count (TC) counts the number of triangles in the graph; and, finally, Single-source Shortest Paths (SSSP) uses $\delta$-stepping~\cite{meyer2003delta} to return the distance of all vertices of a graph to a given source vertex. Table~\ref{tab:graph_kernels} shows the main characteristics of these six applications, including the size of property array elements, and input parameters such as the execution style (push or pull), or the use of frontiers. 

\begin{table}
        \begin{center}
        \small
        \setlength\tabcolsep{6pt}
        \bgroup
        \def\arraystretch{0.7}
    \resizebox{\columnwidth}{!}{%
        \begin{tabular}{@{}ll@{}}
            \textbf{Component}   & \textbf{Description}\\ \midrule
            {\textbf{Branch Predictor}} & hashed-perceptron \\
            \cmidrule(l{.4em}){2-2}
            {\textbf{CPU}} & \SI{3.8}{\giga\hertz}, 4-wide out-of-order processor \\
                           & 6-stage pipeline, 224-entries re-order buffer \\
            \cmidrule(l{.4em}){2-2}
            {\textbf{L1 ITLB}}   & 64-entry, 4-way, 1cc, 8-entry MSHR, LRU \\
            \cmidrule(l{.4em}){2-2}
            {\textbf{L1 DTLB}}   & 64-entry, 4-way, 1cc, 8-entry MSHR, LRU \\
            \cmidrule(l{.4em}){2-2}
            {\textbf{L2 TLB}}    & 1536-entry, 12-way, 8cc, 16-entry MSHR, LRU \\
            \cmidrule(l{.4em}){2-2}
            {\textbf{L1I Cache}} & \SI{32}{\kilo\byte}, 8-way, 4cc, 10-entry MSHR, LRU \\
            \cmidrule(l{.4em}){2-2}
            {\textbf{L1D Cache}} & \SI{32}{\kilo\byte}, 8-way, 4cc, 10-entry MSHR, LRU, IPCP\cite{9138971} or Berti\cite{navarro2022berti}  \\
			\cmidrule(l{.4em}){2-2}
            {\textbf{L2 Cache}} & \SI{1}{\mega\byte}, 16-way, 10cc, 16-entry MSHR, LRU, SPP \cite{kim2016path}  \\
            \cmidrule(l{.4em}){2-2}
            {\textbf{LLC}}    & \SI{1.375}{\mega\byte} per core, 11-way, 36/56cc, 64-entry MSHR, LRU\\
            \cmidrule(l{.4em}){2-2}
                                                & \SI{16}{\giga\byte}, DDR4 SDRAM \\
            \multicolumn{1}{c}{\textbf{DRAM}}   & single-core data-rate: \SI{12.8}{GB/s} per core \\
                                                & multi-core data-rate: \SI{3.2}{GB/s} per core \\
                                                & $t_{RP}=t_{RCD}=t_{CAS}=\;$\SI{24}{cycles} \\
            \midrule
        \end{tabular}
    }
        \egroup
    \end{center}
    \caption{System configuration.}
    \label{table:cpu_config}
\end{table}


For each graph-processing kernel, we consider 6 different input graphs that feature different sizes and distributions of node degrees (\emph{e.g.,} power-law, normal, etc.). Different degree distributions produce different memory access patterns. For instance, when node degrees are distributed following a power-law function, there are a few highly connected graph nodes that yield more data reuse opportunities than vertices with a few connections. Table~\ref{tab:input_graphs} lists all considered input graphs.

In addition, we only consider workloads for which the baseline system shows LLC MPKI greater than 1. This filters out workloads and leaves us with 31 GAP workloads and 24 SPEC workloads.

All workload traces have been obtained using the SimPoint methodology \cite{perelman_using_nodate} to identify at least one SimPoint representative of each workload. Each SimPoint is 1 billion instructions long and characterizes a different phase of these workloads, 
similar to prior work \cite{jimenez_multiperspective_2017,shi_applying_2019,5695535,9499825}.

Section~\ref{sec:evaluation} refers to SPEC 2006 and SPEC 2017 workloads as SPEC and to GAP workloads as GAP.

\begin{table}
    \begin{center}
        \small
        \setlength\tabcolsep{6pt}
        \bgroup
        \def\arraystretch{0.7}
    \resizebox{\columnwidth}{!}{%
        \begin{tabular}{@{}lllllll@{}}
            \midrule
            & \textbf{BC}~\cite{beamer_gap_2015} & \textbf{BFS}~\cite{beamer_gap_2015} & \textbf{CC}~\cite{beamer_gap_2015} & \textbf{PR}~\cite{beamer_gap_2015} & \textbf{TC}~\cite{beamer_gap_2015} & \textbf{SSSP}~\cite{beamer_gap_2015} \\
            \cmidrule(l{.4em}){2-7}
            \textbf{irregData ElemSz} & \SI{8}{\byte} + \SI{4}{\byte} & \SI{4}{\byte} & \SI{4}{\byte} & \SI{4}{\byte} & \SI{4}{\byte} & \SI{4}{\byte} \\
            \cmidrule(l{.4em}){2-7}
            \textbf{Execution style} & Push-Mostly & Push \& Pull & Push-Mostly & Pull-Only & Push-Only & Push-Only \\
            \cmidrule(l{.4em}){2-7}
            \textbf{Use Frontier} & Yes & Yes & No & No & No & Yes \\
            \midrule
        \end{tabular}
    }
        \egroup
    \end{center}
    \caption{Graph kernels}
    \label{tab:graph_kernels}
\end{table}

\begin{table}
    \begin{center}
        \small
        \setlength\tabcolsep{6pt}
        \bgroup
        \def\arraystretch{0.7}
    \resizebox{\columnwidth}{!}{%
    \begin{tabular}{@{}lllllll@{}}
        \midrule
         & \textbf{Web}\cite{beamer_gap_2015} & \textbf{Road}\cite{beamer_gap_2015} & \textbf{Twitter}\cite{beamer_gap_2015} & \textbf{Kron}\cite{beamer_gap_2015} & \textbf{Urand}\cite{beamer_gap_2015} & \textbf{Friendster}\cite{yang_community-affiliation_2012} \\
         \cmidrule(l{.4em}){2-7}
         \textbf{\# Vertices (in M)} & 50.6 & 23.9 & 61.6 & 134.2 & 134.2 & 65.6 \\
         \cmidrule(l{.4em}){2-7}
         \textbf{\# Edges (in M)} & 1,949.4 & 58.3 & 1,468.4 & 2,111.6 & 2,147.4 & 3,612.1 \\
        \midrule
    \end{tabular}
    }
        \egroup
    \end{center}
    \caption{Input Graphs}
    \vspace{-0.5cm}
    \label{tab:input_graphs}
\end{table}

\subsection{Single-Core Evaluation}
\label{subsec:single_core_workloads}

Our set of single-core workloads contains 55 distinct workloads: 31 possible combinations of graph-processing kernels and input graphs, described in Section~\ref{subsec:workloads}, and 24 SPEC CPU 2006 \cite{spec2006} and SPEC CPU 2017~\cite{spec2017} benchmarks. All considered workloads experience at least 1 Miss per Kilo Instructions (MPKI) in the baseline system that Table~\ref{table:cpu_config} describes. Each workload is executed for 100 million instructions to warm up the memory hierarchy and the other microarchitectural structures, and it is executed for an additional set of 100 million instructions to obtain performance data.
We run experiments evaluating the impact of using larger numbers of instructions (500 million warmup instructions, 1 billion simulation instructions), and observe identical trends with negligible differences in terms of IPC.

\subsection{Multi-Core Evaluation}
\label{subsec:multi_core_workloads}

We generate multi-core workload mixes using the same methodology as previous work~\cite{bera_hermes}. We consider either single-core GAP workloads or single-core SPEC workloads 
to create both homogeneous and heterogeneous multi-core workload mixes. To generate the homogeneous ones, we randomly select 50 single-core workloads and run four instances of each workload, one per core. For the heterogeneous mixes, we randomly select 50 combinations of four single-core workloads. In total, we consider 50 homogeneous and 50 heterogeneous four-core workloads. We do this process for both SPEC and GAP benchmark suites, meaning that our multi-core evaluation campaign is composed of 200 workloads. Finally, our multi-core experiments use the same number of warmup and simulation instructions as the single-core scenario.

Our performance results concerning multi-core workloads report the weighted speedup normalized to the baseline. This metric is commonly used to evaluate multi-core workloads~\cite{jimenez_multiperspective_2017, shi_applying_2019,9923823} since it avoids performance overestimation due to high-IPC threads. The metric is computed as follows: for each single-core workload, we compute its IPC in a multi-core scenario shared with the other co-running single-core workloads ($IPC_{shared}$), and its IPC running in isolation on the same system ($IPC_{single}$). We then compute the weighted IPC of the mix as the weighted sum of $IPC_{shared}/IPC_{single}$ for all the benchmarks in the mix, and we normalize this weighted IPC with the weighted IPC of the baseline design.

\subsection{Alternative Techniques}
\label{subsec:alternative_approaches}

Besides \texttt{TLP}, we consider the following techniques in our evaluation: 
(i) the Perceptron-based Prefetch Filtering (\texttt{PPF})~\cite{bhatia2019perceptron}, a perceptron-based predictor that filters inaccurate prefetch requests, thus increasing the accuracy of the underlying prefetcher. PPF is located at the L2C since it is built on top of the SPP prefetcher. When using PPF, we configure SPP as previous work indicates~\cite{bhatia2019perceptron} to fully exploit the advantages of PPF. 
(ii) \texttt{Hermes}~\cite{bera_hermes}, the state-of-the-art off-chip predictor that removes long-lasting load requests from the critical path by issuing speculative requests to the DRAM controller. 
(iii) \texttt{Hermes+PPF}, a scheme that uses both \texttt{Hermes} as off-chip predictor and \texttt{PPF} as a prefetch filter.


\section{Evaluation}
\label{sec:evaluation}









\subsection{Single-Core Evaluation}
\label{subsec:single_core_analysis}

\begin{figure}
    \centering
    \begin{subfigure}[b]{1.0\columnwidth}
        \centering
        \includegraphics[width=\columnwidth]{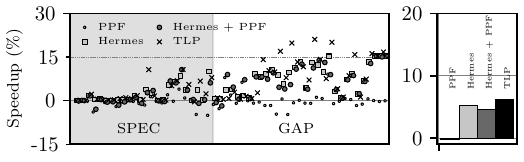}
        \caption{IPCP
        }
        \label{fig:evaluation_single_core_speedup_ipcp}
    \end{subfigure}
    \hfill
    \begin{subfigure}[b]{1.0\columnwidth}
        \centering
        \includegraphics[width=\columnwidth]{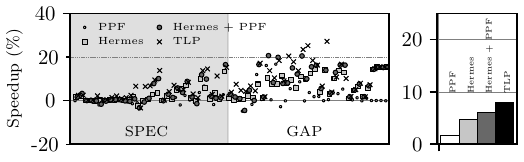}
        \caption{Berti
        }
        \label{fig:evaluation_single_core_speedup_berti}
    \end{subfigure}
    \caption{Performance evaluation in the single-core scenario.  
    }
    \vspace{-0.5cm}
    \label{fig:evaluation_single_core_speedup}
\end{figure}

This section evaluates \texttt{TLP} in the single-core context following the methodology that Section~\ref{sec:methodology} presents. 
Figure~\ref{fig:evaluation_single_core_speedup} shows the performance gains provided by \texttt{PPF}, \texttt{Hermes}, \texttt{Hermes+PPF}, and \texttt{TLP} in the single-core context over a baseline described in Section~\ref{sec:methodology}, which has no off-chip predictor neither L1D prefetch filter. Specifically, Figures~\ref{fig:evaluation_single_core_speedup_ipcp} and~\ref{fig:evaluation_single_core_speedup_berti} present performance results when using IPCP and Berti as L1D prefetchers, respectively. 
Both figures display speedup with respect to the baseline system in the y-axis. The x-axis shows the selected SPEC and GAP workloads. For each benchmark suite we sort workloads in increasing order of MPKI in the baseline system.
This evaluation indicates that \texttt{TLP} significantly outperforms state-of-the-art approaches for off-chip prediction (\texttt{Hermes}), prefetch filtering (\texttt{PPF}), and a combination of them (\texttt{Hermes+PPF}). 
In the scenario considering IPCP as L1D prefetcher, \texttt{TLP} yields 6.2\% geometric mean speedup with respect to the baseline system while \texttt{PPF}, \texttt{Hermes}, and \texttt{Hermes+PPF} bring -0.2\%, 5.2\%, and 4.7\% geometric mean speedups, respectively. 
When considering the Berti prefetcher, \texttt{TLP} yields 8.1\% geometric mean speedup as compared to 1.7\%, 4.8\%, and 6.1\% for \texttt{PPF}, \texttt{Hermes}, and \texttt{Hermes+PPF}, respectively. 
\texttt{TLP} achieves larger performance gains for GAP than SPEC. Since GAP workloads are strongly memory bound, the reductions in terms of DRAM transactions that \texttt{TLP} achieves compared to \texttt{Hermes} particularly benefit GAP.

\begin{figure}
    \centering
    \begin{subfigure}[b]{1.0\columnwidth}
        \centering
        \includegraphics[width=\columnwidth]{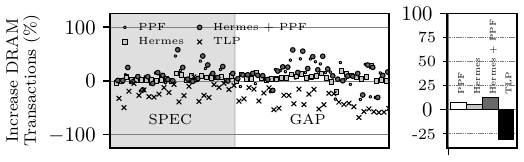}
        \caption{IPCP
        }
        \label{fig:dram_transaction_single_core_ipcp}
    \end{subfigure}
    \hfill
    \begin{subfigure}[b]{1.0\columnwidth}
        \centering
        \includegraphics[width=\columnwidth]{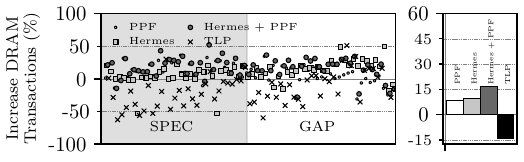}
        \caption{Berti
        }
        \label{fig:dram_transaction_single_core_berti}
    \end{subfigure}
    \caption{Increase in DRAM transactions in the single-core scenario. Lower is better.
    }
    \vspace{-0.5cm}
    \label{fig:dram_transaction_single_core}
\end{figure}

To identify the source of TLP performance improvements we quantify the impact of \texttt{PPF}, \texttt{Hermes}, \texttt{Hermes+PPF}, and \texttt{TLP} in terms of number of DRAM accesses.   Figure~\ref{fig:dram_transaction_single_core} shows this evaluation. 
Specifically, Figure~\ref{fig:dram_transaction_single_core_ipcp} presents results obtained using IPCP, while Figure~\ref{fig:dram_transaction_single_core_berti} considers Berti. 
The y-axis displays the increase in DRAM transactions processed by the memory controller in the single-core context while the x-axis shows the SPEC and GAP workloads sorted in terms of MPKI.
When using IPCP, \texttt{TLP} reduces DRAM transactions by an average of 30.7\% over the baseline while \texttt{PPF}, \texttt{Hermes}, and \texttt{Hermes+PPF} increase average DRAM transactions by 7.7\%, 5.2\%, and 13.3\%, respectively. 
When considering Berti, \texttt{TLP} reduces the number of DRAM transactions by an average of 14.2\% over the baseline, while \texttt{PPF}, \texttt{Hermes}, and \texttt{Hermes+PPF} trigger of 8.8\%, 9.6\%, and 16.9\% additional DRAM transactions, respectively. 

\begin{figure}
    \centering
    \begin{subfigure}[b]{1.0\columnwidth}
        \centering
        \includegraphics[width=\columnwidth]{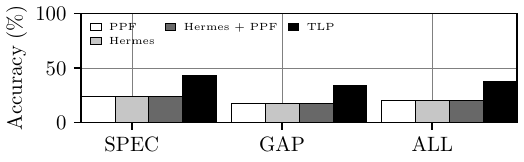}
        \caption{IPCP 
        }
        \label{fig:evaluation_l1d_ipcp_prefetcher_accuracy}
    \end{subfigure}
    \hfill
    \begin{subfigure}[b]{1.0\columnwidth}
        \centering
        \includegraphics[width=\columnwidth]{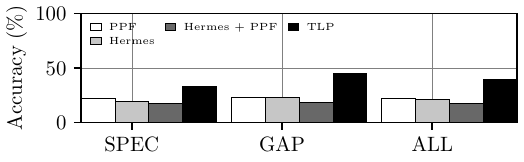}
        \caption{Berti 
        }
        \label{fig:evaluation_l1d_berti_prefetcher_accuracy}
    \end{subfigure}
    \caption{Accuracy of the L1D prefetchers. 
    }
    \vspace{-0.5cm}
    \label{fig:evaluation_l1d_prefetcher_accuracy}
\end{figure}

To further explain the performance gain obtained by \texttt{TLP}, we evaluate the accuracy of the considered L1D prefetchers (IPCP and Berti) when \texttt{PPF}, \texttt{Hermes}, \texttt{Hermes+PPF}, and \texttt{TLP} operate in the system.
Figure~\ref{fig:evaluation_l1d_prefetcher_accuracy} presents the results. Specifically, Figures \ref{fig:evaluation_l1d_ipcp_prefetcher_accuracy} and \ref{fig:evaluation_l1d_berti_prefetcher_accuracy} present the accuracy of the IPCP and Berti prefetchers, respectively. The key takeaway of this comparison is that \texttt{TLP} increases the accuracy of the L1D prefetchers. Across all SPEC and GAP workloads, IPCP experiences an average accuracy of 20.6\%, 20.6\%, 20.3\%, and 38.0\% when \texttt{PPF}, \texttt{Hermes}, \texttt{Hermes+PPF}, and \texttt{TLP} operate in the system, respectively. Finally, we observe similar behavior for the Berti prefetcher; Figure \ref{fig:evaluation_l1d_berti_prefetcher_accuracy} reveals that Berti experiences the highest accuracy with \texttt{TLP}.

Data in Figures~\ref{fig:dram_transaction_single_core} 
 and~\ref{fig:evaluation_l1d_prefetcher_accuracy} indicate that \texttt{TLP} successfully reduces the number of 
DRAM transactions that state-of-the-art off-chip prediction and prefetch filtering approaches trigger. 

\subsection{Multi-Core Evaluation}
\label{subsec:multi_core_analysis}


This section evaluates the performance of \texttt{TLP} in the multi-core scenario following the methodology that Section~\ref{sec:methodology} presents.
In addition, this section 
indicates the contribution of each specific \texttt{TLP} component to final performance 
(Section \ref{subsec:performance_improvement_breakdown}), and evaluates \texttt{TLP} considering different DRAM bandwidth scenarios (Section \ref{subsec:performance_sensitivity_analysis}).  

\begin{figure}
    \centering
    \begin{subfigure}[b]{1.0\columnwidth}
        \centering
        \includegraphics[width=\columnwidth]{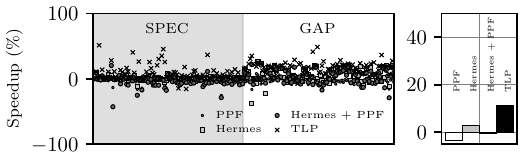}
        \caption{IPCP
        }
        \label{fig:evaluation_multi_core_speedup_ipcp}
    \end{subfigure}
    \hfill
    \begin{subfigure}[b]{1.0\columnwidth}
        \centering
        \includegraphics[width=\columnwidth]{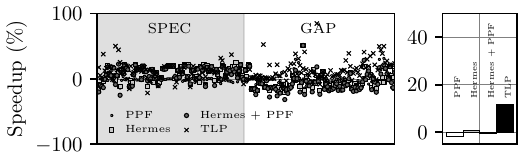}
        \caption{Berti}
        \label{fig:evaluation_multi_core_speedup_berti}
    \end{subfigure}
    \caption{Performance improvement in the multi-core scenario.
    }
    \label{fig:evaluation_multi_core_speedup}
\end{figure}

Figure~\ref{fig:evaluation_multi_core_speedup} 
shows the performance gains provided by \texttt{PPF}, \texttt{Hermes}, \texttt{Hermes+PPF}, and \texttt{TLP} in the multi-core context over the baseline system that Section \ref{sec:methodology} describes. 
Specifically, Figures \ref{fig:evaluation_multi_core_speedup_ipcp} and \ref{fig:evaluation_multi_core_speedup_berti}  present performance results considering IPCP and Berti, respectively. 
Both figures display speedup over the baseline system in the y-axis. The x-axis shows the multi-core SPEC and GAP workloads sorted in increasing order in terms of MPKI. The sorting is done independently within each benchmark suite.
Considering the IPCP prefetcher, \texttt{TLP} improves geometric mean performance by 11.5\% as compared to -3.3\%, 3.0\%, and -0.5\% for \texttt{PPF}, \texttt{Hermes}, and \texttt{Hermes+PPF}, respectively. When considering Berti as L1D prefetcher we observe similar trends. Specifically, \texttt{TLP} yields a 11.8\% geometric mean speedup over the baseline as compared to -1.5\%, 1.0\%, and -0.3\% for \texttt{PPF}, \texttt{Hermes}, and \texttt{Hermes+PPF}, respectively. The main takeaway of this experiment is that \texttt{TLP} provides significantly higher multi-core performance than all considered prior proposals. 

\begin{figure}
    \centering
    \begin{subfigure}[b]{1.0\columnwidth}
        \centering
        \includegraphics[width=\columnwidth]{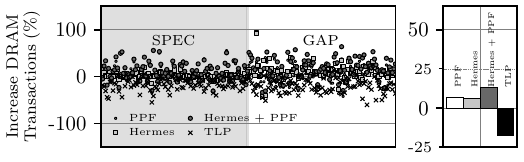}
        \caption{IPCP
        }
        \label{fig:dram_transactions_multi_core_ipcp}
    \end{subfigure}
    \hfill
    \begin{subfigure}[b]{1.0\columnwidth}
        \centering
        \includegraphics[width=\columnwidth]{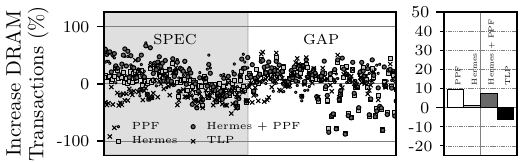}
        \caption{Berti}
        \label{fig:dram_transactions_multi_core_berti}
    \end{subfigure}
    \caption{Increase in DRAM transactions in the multi-core scenario. Lower is better.
    } 
    \vspace{-0.5cm}
    \label{fig:dram_transactions_multi_core}
\end{figure}

To explain the source of \texttt{TLP} performance improvements in the multi-core context, we quantify the impact of \texttt{PPF}, \texttt{Hermes}, \texttt{Hermes+PPF}, and \texttt{TLP} on the number of DRAM accesses.
Figure~\ref{fig:dram_transactions_multi_core} shows the increase in terms of DRAM transactions over the  baseline system that Section~\ref{sec:methodology} describes. 
Figures 
\ref{fig:dram_transactions_multi_core_ipcp} and 
Figure~\ref{fig:dram_transactions_multi_core_berti} show the impact on DRAM transactions when IPCP and Berti operate at L1D, respectively.  
Considering the IPCP prefetcher, \texttt{TLP} reduces the number of DRAM transactions by an average of 17.7\% over the baseline while \texttt{PPF}, \texttt{Hermes}, and \texttt{Hermes+PPF} increase DRAM transactions by 6.5\%, 6.0\%, and 13.4\%, respectively. 
We observe a similar behavior when Berti is used as L1D prefetcher: \texttt{TLP} reduces the average DRAM transactions by 6.3\% while \texttt{PPF}, \texttt{Hermes}, and \texttt{Hermes+PPF} increase DRAM transactions by 9.8\%, 1.4\%, and 7.8\%, respectively. The main takeaway is that \texttt{TLP} outperforms prior approaches for off-chip prediction and prefetch filtering in multi-core contexts since it significantly reduces the DRAM pressure. 

\subsubsection{Performance Contribution of Each TLP Component}
\label{subsec:performance_improvement_breakdown}



This section evaluates the contribution of each specific \texttt{TLP} component to the final performance.
To do so, we consider five different scenarios besides \texttt{TLP}: 
i) \texttt{FLP}, which consits of just the FLP predictor without the selective delay mechanism. ii) \texttt{SLP}, which consists of just the SLP predictor.
iii) \texttt{Two-Step Predictor (TSP)}, which consists of FLP without the selective delay mechanism, and SLP without the feature based on FLP output. \texttt{TSP} consumes FLP predictions before 
the completion of L1D accesses
, as \texttt{Hermes} does.
Therefore, the difference between \texttt{TSP} and \texttt{Hermes} is the use of SLP.
iv) \texttt{Delayed TSP}, a technique similar to TSP with the exception that always delays the consumption of FLP predictions upon L1D misses. 
v) \texttt{Selective TSP}, an evolution of \texttt{Delayed TSP} that uses selective delay. Finally, we consider \texttt{TLP}.
The difference between \texttt{TLP} and \texttt{Selective TSP} is that \texttt{TLP} uses a feature based on the output of FLP to drive the predictions of SLP.
\begin{figure}
    \centering
    \includegraphics{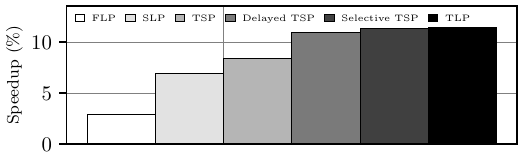}
    \caption{
    Performance contribution of each \texttt{TLP} component.
    }
    \vspace{-0.5cm}
    \label{fig:evaluation_multicore_performance_breakdown}
\end{figure}
Figure~\ref{fig:evaluation_multicore_performance_breakdown} shows the performance of these 
six
approaches when IPCP operates at L1D. We observe that by incrementally adding parts of our design to form our final proposal, \texttt{TLP}, we can compound performance as 
\texttt{FLP}, \texttt{SLP},
\texttt{TSP}, \texttt{Delayed TSP}, and \texttt{Selective TSP} yield respectively 
2.9\%, 6.9\%,
8.4\%, 10.2\%, and 11.4\% geometric mean 
speedups
over the baseline. Our final proposal, \texttt{TLP} provides a 11.5\% speedup over the baseline, justifying our design choices. 
Although the difference between \texttt{TLP} and \texttt{Selective TSP} in terms of geometric mean speed-ups is rather small, it becomes larger for workloads with 
high correlation between off-chip load demand requests and off-chip L1D prefetch requests
like \texttt{bc.road}.
We observe a very similar behavior when we consider Berti as L1D prefetcher. 

\begin{figure}
    \centering
    \begin{subfigure}[b]{1.0\columnwidth}
        \centering
        \includegraphics[width=\columnwidth]    {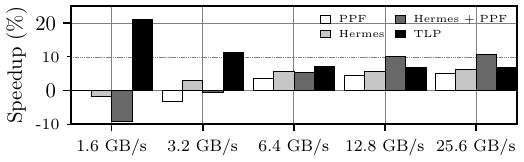}
        \caption{Performance improvement.}
        \label{fig:tlp_performance_against_dram_bw}
    \end{subfigure}
    \hfill
    \begin{subfigure}[b]{1.0\columnwidth}
        \centering
        \includegraphics[width=\columnwidth]{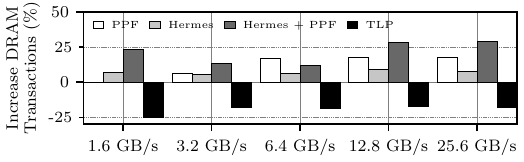} 
        \caption{Increase in DRAM transactions. 
        }
        \label{fig:tlp_dram_trans_against_dram_bw}
    \end{subfigure}
    \caption{Impact of DRAM bandwidth on geometric mean  performance and average number of DRAM transactions in the multi-core context.
    }
     \vspace{-0.5cm}
    \label{fig:tlp_against_dram_bw}
\end{figure}

\subsubsection{Sensitivity Analysis on DRAM Bandwidth}
\label{subsec:performance_sensitivity_analysis}

This section evaluates \texttt{TLP}, \texttt{PPF}, \texttt{Hermes}, and \texttt{Hermes+PPF} on scenarios with 1.6 GB/s per core up to 25.6 GB/s per core. Figures~\ref{fig:tlp_performance_against_dram_bw} and~\ref{fig:tlp_dram_trans_against_dram_bw} show the geometric mean performance and the impact on DRAM accesses of \texttt{TLP} of these four approaches in the multi-core context, respectively.  

Figure~\ref{fig:tlp_performance_against_dram_bw} indicates that \texttt{TLP} outperforms the other approaches under memory bandwidth regimes between 1.6 and 6.4 GB/s per core.
The improvement achieved by \texttt{TLP} in the 1.6 GB/s scenario is 21.2\%, while \texttt{TLP} obtains a 6.9\% geometric mean speedup over the baseline when there are 25.6 GB/s per core available.
Even in scenarios with unrealistically large memory bandwidth per core ratios (\textit{e.g.}, 12.8 or 25.6 GB/s per core), \texttt{TLP} outperforms \texttt{Hermes} and \texttt{PPF} since it avoids cache pollution due to inaccurate prefetching. 
In these scenarios, the unrealistic aboundance of memory bandwidth allows \texttt{Hermes+TLP} to deliver larger performance than \texttt{TLP}.



Figure~\ref{fig:tlp_dram_trans_against_dram_bw} shows the impact in terms of DRAM transactions for all considered approaches on five memory bandwidth per core scenarios. 
\texttt{TLP} achieves a remarkable reduction in terms of DRAM transactions over the baseline in all scenarios.
Specifically, \texttt{TLP} decreases DRAM transactions from 24.8\% (1.6 GB/s core scenario) to 17.6\% (25.6 GB/s per core scenario) as compared to the baseline.
\subsection{Designs Enhanced with TLP's Storage Budget}
\label{subsec:additional_design_comparisons}



\begin{figure}
    \centering
    \begin{subfigure}[b]{1.0\columnwidth}
        \centering
        \includegraphics[width=\columnwidth]    {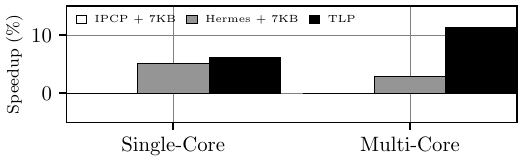}
        \caption{
        IPCP
        }
        \label{fig:}
    \end{subfigure}
    \hfill
    \begin{subfigure}[b]{1.0\columnwidth}
        \centering
        \includegraphics[width=\columnwidth]{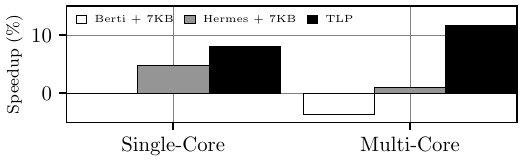} 
        \caption{
        Berti
        }
        \label{fig:}
    \end{subfigure}
    \caption{
    Performance improvement of designs enhanced with TLP's storage budget.
     \vspace{-0.5cm}
    }
    \label{fig:iso_design_comparisons}
\end{figure}

In addition to the results provided in the previous sections, we also evaluate other designs leveraging 7KB of extra storage over IPCP, Berti, and \texttt{Hermes}. We compare them to \texttt{TLP}. Figure~\ref{fig:iso_design_comparisons} shows the evaluation of these designs in both single-core and multi-core contexts.
In the single-core context, adding 7KB of extra storage to IPCP and Berti does not leverage any performance benefits over the baseline. When Hermes is enhanced with 7KB of extra storage, it provides performance improvements close to the baseline Hermes, $i.e.$, 5.2\% and 4.8\% geometric mean speedup for IPCP and Berti, respectively. In comparison, \texttt{TLP} leverages 6.2\% and 8.1\% geometric mean speedup for IPCP and Berti.
In the multi-core context, we observe a similar behavior where adding extra storage to the prefetchers does not leverage performance improvements. Finally, we observe that Hermes shows a similar behavior as its counterpart using no extra storage.

\section{Related Work}
\label{sec:related_work}

To the best of our knowledge, this is the first work to provide a cooperative solution for off-chip prediction and adaptive prefetch filtering using neural methods. Sections \ref{sec:background_motivation}, \ref{subsec:impact_hermes}, and \ref{sec:evaluation} describe, analyze and compare our proposal against Hermes~\cite{bera_hermes}, the state-of-the-art off-chip predictor, respectively. Sections \ref{sec:background_motivation} and \ref{sec:evaluation} compare our proposal against PPF~\cite{bhatia2019perceptron}, the state-of-the-art prefetch filter. This section focuses on other related work targeting memory hierarchy optimizations.

\vspace{0.1cm}
\emph{\textbf{Hit/Miss Prediction.}} 
Jalili and Erez~\cite{jalili_locmap} proposed \textit{LP}, a scheme that uses a flat-array to track the residency of cache lines in the cache hierarchy. This flat-array is stored in a reserved section of DRAM, and a small cache keeps recently used entries of the flat-array for future predictions. \textit{LP} presents several challenges. First, it can have a high false-positive prediction rate. Second, the size of \textit{LP} can grow very large, leading to significant latency and storage overheads. Third, \textit{LP} does not address the large bandwidth consumption of cache prefetchers. In contrast, TLP requires only 4KB per core of storage overhead, and 56 additional bits per prefetch request, while producing accurate off-chip predictions for both load and prefetch requests. With its more streamlined approach, TLP offers a promising solution to the challenges presented by other prefetching methods, paving the way for improved performance and efficiency in modern computing systems. 

\vspace{0.1cm}
\emph{\textbf{Data Prefetching.}} Stream and strided cache prefetchers are unable to effectively prefetch for the indirect memory access patterns of graph-processing workloads \cite{basak_analysis_2019,balaji_p-opt_nodate}. Yu et al. \cite{yu_imp_2015} propose a microarchitectural prefetcher that identifies and prefetches indirect memory access patterns without requiring any application nor software information. Ainsworth et al. \cite{10.1145/2925426.2926254} propose a prefetcher that leverages application-level information to capture indirect memory access patterns. Basak et al. \cite{basak_analysis_2019} propose DROPLET, a physically decoupled prefetcher that takes into account the reuse distances when applying prefetching for different graph types. Although effective, these hardware prefetchers increase memory bandwidth consumption. In  contrast, our proposal reduces the cost of hardware prefetching while keeping its advantages, as Section \ref{sec:evaluation} shows. 


\vspace{0.1cm}
\emph{\textbf{Cache Bypassing.}} Recent research has proposed several complex cache replacement and bypassing policies~\cite{jaleel_high_2010,wu2011ship,jimenez_multiperspective_2017,shi_applying_2019}, that have demonstrated significant performance gains in general-purpose computing applications. However, recent studies~\cite{jamet_characterizing_2020} show that these policies are ineffective when applied to workloads managing irregular and sparse structures like graphs due to the irregularity of the memory access patterns that these workloads display.

\vspace{0.1cm}
\emph{\textbf{Memory Optimizations for Graph-Processing Applications.}}
Recent work demonstrates the benefits of optimizing the memory hierarchy for graph applications. Ozdal et al. \cite{7551391} use scratchpads to store vertex and edge data of graph-processing applications, while Gonzalez et al. \cite{10.5555/2685048.2685096} employ a large eDRAM scratchpad to accommodate larger volumes of graph data than the conventional SRAMs. Several prior works \cite{8327036,8327035,7920847,ppa,7284059} reduce the latency cost of graph memory accesses by executing graph-processing operations close to DRAM, partially hiding the latency cost of the corresponding memory accesses. TLP complements these works since it improves the cache management of a wide range of applications.

\vspace{0.1cm}
\emph{\textbf{Redesigning the Cache Hierarchy.}}
The Distill Cache \cite{qureshi_line_2007} approach reserves a section of the L2C to place the used words of a cache line when that line is elected for eviction. This design improves the use of cache storage capacity by evicting just the unused words of each cache line. In contrast, our proposal manages the pervasiveness of highly irregular access patterns and dynamically classifies memory accesses as either regular or irregular. By labeling some memory accesses as not cache-friendly, we avoid cache pollution and useless cache look-ups.
The Victim Cache~\cite{jouppi1990improving} proposal is a small fully-associative cache,  found on the refill path of the LLC. It contains eviction victims of the cache to which it is attached and tries to decrease conflict misses. On LLC misses, both the LLC and the Victim Cache are looked-up; if the requested cache block is found in the Victim Cache, the LLC victim and the Victim Cache entry are swapped, thus lowering the miss latency. On a Victim Cache miss, the block is fetched from DRAM and the LLC victim is inserted in the Victim Cache. While the Victim Cache has been proven effective at improving the performance of SPEC workload \cite{spec2006,spec2017}, it relies heavily on spatial locality as it inserts caches victims. Our proposal does not rely on locality assumptions and shortcuts the cache hierarchy when it is predicted to be inefficient.
\section{Conclusions}
\label{sec:conclusion}

This work introduces \emph{Two Level Perceptron (TLP)} predictor, a neural approach that leverages two perceptron predictors to apply off-chip prediction to both demand and prefetch requests. This technique prevents memory bandwidth waste due to inaccurate prefetch requests or wrong off-chip prediction, and avoids cache pollution due to prefetching. 
We evaluate TLP against several previous approaches (PPF~\cite{bhatia2019perceptron} and Hermes~\cite{bera_hermes}) considering 55 single-core workloads, 200 multi-core workloads, and two state-of-the-art L1D prefetchers, IPCP~\cite{9138971} and Berti~\cite{navarro2022berti}. TLP achieves a 11.5\% geometric mean speedup when deployed on a multi-core system using the IPCP L1D prefetcher, while the best previous approach, Hermes, delivers 3.0\%.
When considering Berti, Hermes delivers 0.9\% geometric mean speedup while TLP obtains 11.8\% improvement.
Our evaluation also demonstrates that TLP significantly reduces the overhead in terms of DRAM bandwidth transactions that previous approaches incur in all considered scenarios.


\ifdefined\hpcacameraready
    \section*{Acknowledgments}
    The authors are grateful to the anonymous MICRO 2023 and HPCA 2024 reviewers for their valuable comments and constructive feedback that significantly improved the quality of the paper. 
    This work has been partially supported by the European HiPEAC Network of Excellence, by the Spanish Ministry of Science and Innovation MCIN/AEI/10.13039/501100011033 (contracts PID2019-107255GB-C21 and PID2019-105660RB-C22) and by the Generalitat de Catalunya (contract 2021-SGR-00763).
    This work is supported by the National Science Foundation through grant CCF-1912617 and generous gifts from Intel.
    Marc Casas has been partially supported by the Grant RYC-2017-23269 funded by MCIN/AEI/10.13039/501100011033 and by ESF Investing in your future.
    Els autors agraeixen el suport del Departament de Recerca i Universitats de la Generalitat de Catalunya al Grup de Recerca "Performance understanding, analysis, and simulation/emulation of novel architectures" (Codi: 2021 SGR 00865).
\fi

\appendix

\section{Artifact Appendix}

\subsection{Abstract}

Our artifact provides i) the implementation of TLP, ii) the simulation infrastructure, iii) the set of workloads, iv) scripts for launching the experiments, and v) Python scripts bundled in Jupyter notebooks to exploit the simulation results and reproduce some of the key figures of this paper.

\subsection{Artifact check-list (meta-information)}


{\small
\begin{itemize}
  \item {\bf Program: } Memory traces of SPEC 2006~\cite{spec2006}, SPEC 2017~\cite{spec2017}, and GAP~\cite{beamer_gap_2015} workloads.
  \item {\bf Compilation: } GNU GCC and CMake.
  \item {\bf Metrics: } Performance improvements, reduction in DRAM transactions, statistics on inaccurate off-chip predictions, L1D useful \& useless prefetches, and L1D prefetchers' accuracy.
  \item {\bf Output: } We provide scripts that generate all single-core figures (Figures~\ref{fig:baseline_mpkis}, \ref{fig:hermes_single_core_motivation}, \ref{fig:hermes_offchip_pred_l1d_hit}, \ref{fig:l1d_useless_prefetches_breakdown}, \ref{fig:l1d_useful_prefetches_breakdown}, \ref{fig:evaluation_single_core_speedup}, \ref{fig:dram_transaction_single_core}, \ref{fig:evaluation_l1d_prefetcher_accuracy}).
  \item {\bf Experiments: } We provide scripts that submit the required jobs. The only requirement is a SLURM manager.
  \item {\bf How much disk space required (approximately)?: } 140GB.
  \item {\bf How much time is needed to prepare workflow (approximately)?: } About 1 hour.
  \item {\bf How much time is needed to complete experiments (approximately)?: } About 12 hours.
  \item {\bf Publicly available?: } Yes.
  \item {\bf Code licenses (if publicly available)?: }
  \item {\bf Workflow framework used?: } SLURM for job management.
  \item {\bf Archived (provide DOI)?: } The code is available at \href{https://doi.org/10.5281/zenodo.10100304}{https://doi.org/10.5281/zenodo.10100304}. The trace set is available in 3 volumes:
    \begin{itemize}
        \item Volume 1: \href{https://doi.org/10.5281/zenodo.10083542}{https://doi.org/10.5281/zenodo.10083542}
        \item Volume 2: \href{https://doi.org/10.5281/zenodo.10088347}{https://doi.org/10.5281/zenodo.10088347}
        \item Volume 3: \href{https://doi.org/10.5281/zenodo.10088525}{https://doi.org/10.5281/zenodo.10088525}
    \end{itemize}
\end{itemize}
}

\subsection{Description}

\subsubsection{How to access}

Our artifact is available at \href{https://doi.org/10.5281/zenodo.10100304}{https://doi.org/10.5281/zenodo.10100304}.

\subsubsection{Hardware dependencies} Any hardware capable of compiling and running ChampSim~\cite{champsim_champsim/champsim_2019}.

\subsubsection{Software dependencies} Our artifact depends on the following tools: CMake, Jupyter, Python 3.8.10, matplotlib, and SLURM.

\subsubsection{Data sets} Memory traces of SPEC 2006~\cite{spec2006}, SPEC 2017~\cite{spec2017}, and GAP~\cite{beamer_gap_2015} workloads.


\subsection{Installation}

First, Download the artifact from our \href{https://github.com/itisntalex/TLP-HPCA30-artifact}{GitHub repository} using the appropriate \texttt{git clone} command. 
Second, download the trace set from the following Zenodo records:

\begin{itemize}
    \item Volume 1: \href{https://doi.org/10.5281/zenodo.10083542}{https://doi.org/10.5281/zenodo.10083542}
    \item Volume 2: \href{https://doi.org/10.5281/zenodo.10088347}{https://doi.org/10.5281/zenodo.10088347}
    \item Volume 3: \href{https://doi.org/10.5281/zenodo.10088525}{https://doi.org/10.5281/zenodo.10088525}
\end{itemize}

\subsection{Experiment workflow}

To reproduce all the single-core figures of this work, take the following steps:

\begin{itemize}
    \item \texttt{cd TLP-HPCA30-artifact}
    \item Move the three volumes of the traces artifact to the root of the code's artifact.
    \item Extract the traces using \texttt{tar -xMf TLP-HPCA30-artifact-traces.VOLUME1.tar}. This command is interactive and will request you to provide the name of the next archive's volume as follows \texttt{TLP-HPCA30-artifact-traces.VOLUME2.tar}, etc. A new directory named \texttt{traces/} will be present in the artifact directory.
    \item set paths and username in \texttt{scripts/\-run\_single\_core.sh} (lines 4, 7, 8, 10, and 11), \texttt{scripts/run\_single\_core\_legacy.sh} (lines 4, 7, 8, 10, and 11), \texttt{scripts/run\_single\_core.job} (lines 6, 9, 10, and 12).
    \item Execute \texttt{./scripts/compile\_single\_core.sh} to compile the binaries.
    \item Execute \texttt{scripts/run\_single\_core.sh} and \texttt{scripts/run\_single\_core\_legacy.sh} to launch all single-core simulations.
\end{itemize}

Running all the jobs takes around 12 hours, depending on the cluster and the number of jobs that can be launched in parallel.

\subsection{Evaluation and expected results}

When all jobs are finished, generate the single-core figures using the Jupyter notebooks provided in the \texttt{notebooks} directory. We recommend using the Jupyer extension in the VS Code editor, as it is how the workflow was originally designed.

The single-core figures will be available in the \texttt{plots/} directory.



\subsection{Methodology}

Submission, reviewing and badging methodology:

\begin{itemize}
  \item \url{https://www.acm.org/publications/policies/artifact-review-badging}
  \item \url{http://cTuning.org/ae/submission-20201122.html}
  \item \url{http://cTuning.org/ae/reviewing-20201122.html}
\end{itemize}


\bibliographystyle{IEEEtranS}
\bibliography{refs}

\end{document}